# Floating nematic phase in colloidal platelet-sphere mixtures

Daniel de las Heras[1], Nisha Doshi[2], Terence Cosgrove[2], Jonathan Phipps[3], David I. Gittins[4], Jeroen S. van Duijneveldt[2] & Matthias Schmidt[1]

[1]Theoretische Physik II, Physikalisches Institut, Universität Bayreuth, D-95440 Bayreuth, Germany, [2]School of Chemistry, University of Bristol, Cantock's Close, Bristol BS8 1TS, UK, [3]Imerys Minerals Ltd, Par Moor Center, Par Moor Road, Par Cornwall PL24 2SQ, UK, [4]Imerys Performance & Filtration Minerals, 1732 N 1st Street, San Jose, CA 95112, USA.

The phase behaviour of colloidal dispersions is interesting for fundamental reasons and for technological applications such as photonic crystals and electronic paper. Sedimentation, which in everyday life is relevant from blood analysis to the shelf life of paint, is a means to determine phase boundaries by observing distinct layers in samples that are in sedimentation-diffusion equilibrium. However, disentangling the effects due to interparticle interactions, which generate the bulk phase diagram, from those due to gravity is a complex task. Here we show that a line in the space of chemical potentials $\mu_i$, where $i$ labels the species, represents a sedimented sample and that each crossing of this sedimentation path with a binodal generates an interface under gravity. Complex phase stacks can result, such as the sandwich of a floating nematic layer between top and bottom isotropic phases that we observed in a mixture of silica spheres and gibbsite platelets.

T he analysis of the effect of gravity on colloidal dispersions dates back to Perrin[1] and is used in computer simulations[2,3] and experiment with e.g. depolarized light scattering[4–6] to obtain the osmotic equation of state over a wide range of densities from a single sample. Added depletion agents, e.g. non-adsorbing polymers[7], modify the effective interactions between particles of the primary component, but are typically gravity-neutral. However, in colloidal mixtures all components are subject to gravity, and hence compete to minimize their gravitational energy. The strength of gravity can be quantified by the gravitational lengths $\xi_i = k_B T/(m_i g)$, where $k_B$ is the Boltzmann constant, $T$ is absolute temperature, $m_i$ is the buoyant mass of species $i$ (obtained by subtracting the solvent background), and $g$ is the acceleration due to gravity. Colloidal liquid crystals[8–11] constitute ideal candidates for the investigation of gravitational effects due to their rich phase behaviour.

We used electro-sterically stabilized mixtures of gibbsite platelets and alumina-coated silica spheres (Klebosol 30CAL25 and 30CAL50). Particle dimensions were obtained from transmission electron microscopy and atomic force microscopy (AFM). The average bare diameter of the spheres was $\sigma_S = 30$ nm and 74 nm for Klebosol 30CAL25 and 30CAL50, respectively, with polydispersity of ca. 15% and 21% (3 volume % of the particles are below 60 nm in the case of 30CAL50). The average diameter of the platelets was 186 nm with 29% polydispersity. AFM measurements of a diluted platelet sample deposited on a mica substrate gave a bare platelet thickness of $d \approx 5$ nm. As such results can easily be affected by differences in chemistry between substrate and platelets, we rather treat $d$ as an adjustable parameter, with $d = 3.7$ nm giving the best agreement of the isotropic-nematic (IN) coexistence densities obtained in theory[12] and experiment[13]. The value of $d$ enters the conversion from packing fraction, $\eta_i$, to number density, $\rho_i$, via $\rho_i = \eta_i/v_i$, where $v_i$ is the bare particle volume of species $i = S$ (spheres), $P$ (platelets). Both species were purified by dialysis against deionised water containing 5 mM NaCl in order to screen electrostatic interactions. Both species had a stabilizer (Solsperse 41000) adsorbed onto them. As a result, small-angle neutron scattering showed effective particle diameters $\sigma_i^*$ that were larger by $\approx 10$ nm than the bare sizes, hence $\sigma_i^* = \sigma_i + 10$ nm. The sample preparation follows that in a closely related system[13]. The (bare) packing fraction of spheres was $\eta_S = 0.05$ throughout, and we considered (bare) platelet packing fractions $\eta_P = 0.01, 0.025$ and 0.05. The gravitational lengths were $\xi_P = 2.92$ mm for the platelets, $\xi_S = 22.4$ mm for the small spheres, and $\xi_S = 1.49$ mm for the large spheres. These were obtained by considering the mass density for silica 2.30 g/cm³, for gibbsite 2.42 g/cm³, and for the aqueous solvent 1.00 g/cm³. The density of the adsorbed layer of stabilizer is 1.02 g/cm³. As the latter value is very close to the solvent density, we neglected its effect on the buoyant masses and hence the gravitational lengths of both species. Experiments were carried out at room temperature, $T = 293$ K. The value of the temperature is irrelevant for the slope of the sedimentation path, as this depends only the ratio of the gravitational lengths, such that the temperature dependence of the $\xi_i$ cancels out.

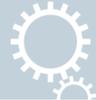

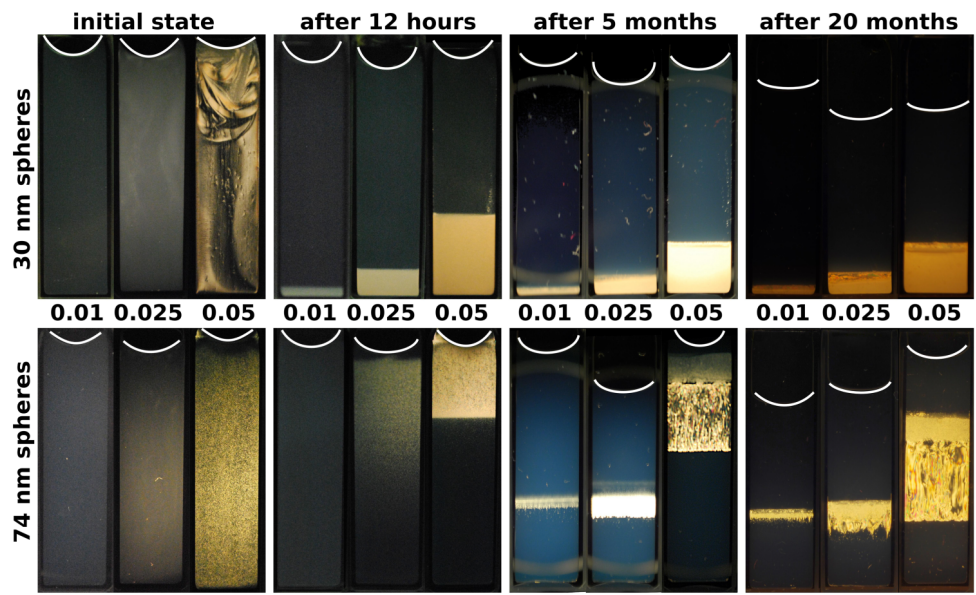

**Figure 1** | Photographs taken with white-light illumination of samples between crossed polarizers, indicating the time evolution of sedimentation in colloidal mixtures of gibbsite platelets with diameter $\sigma_P = 186$ nm and silica spheres with diameter $\sigma_S = 30$ nm (top row) and 74 nm (bottom row). The initial bare packing fraction of spheres was $\eta_S = 0.05$, and of platelets $\eta_P = 0.01$ (left), 0.025 (middle), and 0.05 (right, as indicated). Both sets of three samples are shown in the initial state immediately after filling (first column), after 12 hours (second column), after 5 months (third column), and after 20 months (fourth column) under gravity. The internal width of the vials was 10 mm, the height 40 mm, and the thickness 1 mm. The position of the meniscus between the dispersion and the air is highlighted (white lines). The blurred arcs in the lower part of the pictures for $\eta_S = 0.01$ and 0.025 (bottom row) are artifacts due to the photographic setup.

## Results

Photographs of the time evolution of the different mixtures are presented in Fig. 1. The initial filling process effectively homogenized the systems, with nearly uniform birefringence indicating that nematic order formed quickly at short times. After 12 hours the mixtures with small spheres had separated into a lower nematic and a supernatant isotropic phase. Subsequently, the interface moved downwards compressing the nematic phase. The time evolution of the mixtures with the larger spheres was surprisingly different. While the sample with $\eta_P = 0.01$ remained nearly isotropic after 12 hours, that with $\eta_P = 0.025$ had developed a weak gradient in birefringence. Only the sample with $\eta_P = 0.05$ showed clear phase separation, albeit in reversed order, with a top nematic and bottom isotropic layer. After 5 months its IN interface had moved to smaller heights. The two other samples, however, had developed a floating nematic layer between isotropic bottom and top layers. This state remained stable and could be observed after 20 months. Even the sample with $\eta_P = 0.05$ had transformed into a thick floating nematic.

In order to rationalize these observations, we use a density functional theory for a mixture of hard spheres and vanishingly thin hard platelets[14] with (effective) diameters $\sigma_i^*$. The (Helmholtz) free energy functional, $F[\rho_S, \rho_P]$, reduces to the Rosenfeld functional[15] for pure hard spheres, and includes contributions up to third-order in platelet density; here $\rho_i$ is the one-body density distribution of species $i$. The theory goes beyond Onsager's treatment[16], and accurately describes the IN phase transition in pure hard platelets[12]. Numerically solving the Euler-Lagrange equations, $\delta F[\rho_S, \rho_P]/\delta \rho_i = \mu_i$, gives the equilibrium densities $\rho_i^{\text{eq}}(\mu_S, \mu_P)$, for $i = S, P$.

We use the common splitting of the free energy functional into ideal and excess contributions, $F = F_{\text{id}} + F_{\text{exc}}$. Here the ideal free energy functional for the current binary system is $F_{\text{id}} = \int d\mathbf{r} \rho_S(\mathbf{r}) \left[\ln\left(\rho_S(\mathbf{r})\Lambda_S^3\right) - 1\right] + \int d\mathbf{r} \int d\boldsymbol{\omega} \rho_P(\mathbf{r},\boldsymbol{\omega}) \left[\ln\left(\rho_P(\mathbf{r},\boldsymbol{\omega})\Lambda_P^3\right) - 1\right]$, where $\mathbf{r}$ is the position coordinate and $\boldsymbol{\omega}$ is a unit vector normal to the platelet surface that describes the platelet orientation; the spatial integral is over the system volume, and the orientation integral is over the unit sphere; $\Lambda_i$ is the (irrelevant) de Broglie wavelength of species $i = S, D$. The excess free energy functional $F_{\text{exc}}$ involves convolutions of both bare density fields with appropriate weight functions. The resulting weighted densities are combined into a free energy density per volume, and integrating this over the system volume yields the excess free energy. For further details of $F_{\text{exc}}$ we refer the reader to the original publication devoted to a ternary sphere-needle-platelet mixture[14] and to the recent description of the platelet-sphere subsystem[17], where the binary platelet-sphere functional is described in detail.

Equating pressure, $T$, and $\mu_i$ in both phases gives the bulk phase diagram, which we show in the density representation in Fig. 2a (b) for the mixture with smaller (larger) spheres. The topology of the phase diagram is the same in both systems. Pure hard platelets have a very weak first-order phase transition that arises from the competition of rotational and translational contributions to the free energy: in the nematic phase the loss of orientational entropy is over-compensated by the more efficient packing of the aligned platelets[18,19]. Adding spheres widens the density gap considerably, and these prefer the isotropic phase over the nematic phase. For high densities an isotropic state, almost pure in spheres, coexists with a nematic phase that is almost pure in platelets.

We expect that equilibration on small length scales had occurred in experiment after 12 hours, and hence represent the average densities as statepoints. For the statepoints inside the biphasic region, the relative position on the corresponding tieline directly gives the relative volumes of the two phases, cf. the cartoons in Fig. 2a and b. The top nematic phase in Fig. 2b is a simple consequence of its smaller (in this case) total mass density, $m_S \rho_S + m_P \rho_P$. Besides the thin nematic bottom layer for the small spheres at $\eta_P = 0.01$ and the birefringence gradient in the large sphere system for $\eta_P = 0.025$, the agreement of theoretical and experimental results is remarkable.

In order to analyze the late stages of equilibration, we turn to the chemical potential representation of the phase diagram, cf. Fig. 2c (d) for the mixture with small (large) spheres. The biphasic region collapses to a binodal line. The binodals of both systems are again very similar to each other, with a vertical slope for pure platelets, $\mu_S \rightarrow$ $-\infty$, strong negative curvature upon increasing $\mu_S$ and subsequent point of inflection. The grand potential of the inhomogeneous system can be decomposed[20,21] as $\Omega = \mathcal{U} - T\mathcal{S} + a\sum_i \int_0^H dz \rho_i(z)\phi_{ext,i}(z) - \sum_i \mu_i N_i$, where $\mathcal{U}$ is the internal energy, $\mathcal{S}$ is the entropy, $a$ is the transverse system area, $\phi_{ext,i}(z) = m_i g z$ is the external potential, $z$ is the height coordinate, $H$ is the sample height, and $N_i = a\int_0^H dz \rho_i(z)$ is the total number of particles of species $i$. Hence $\Omega = \mathcal{U} - T\mathcal{S} - a\sum_i \int_0^H dz \rho_i(z) \mu_i(z)$, where we combined the bulk chemical potential and the gravitational energy into a local chemical potential,

$$\mu_i(z) = \mu_i - k_B T z/\xi_i. \quad (1)$$

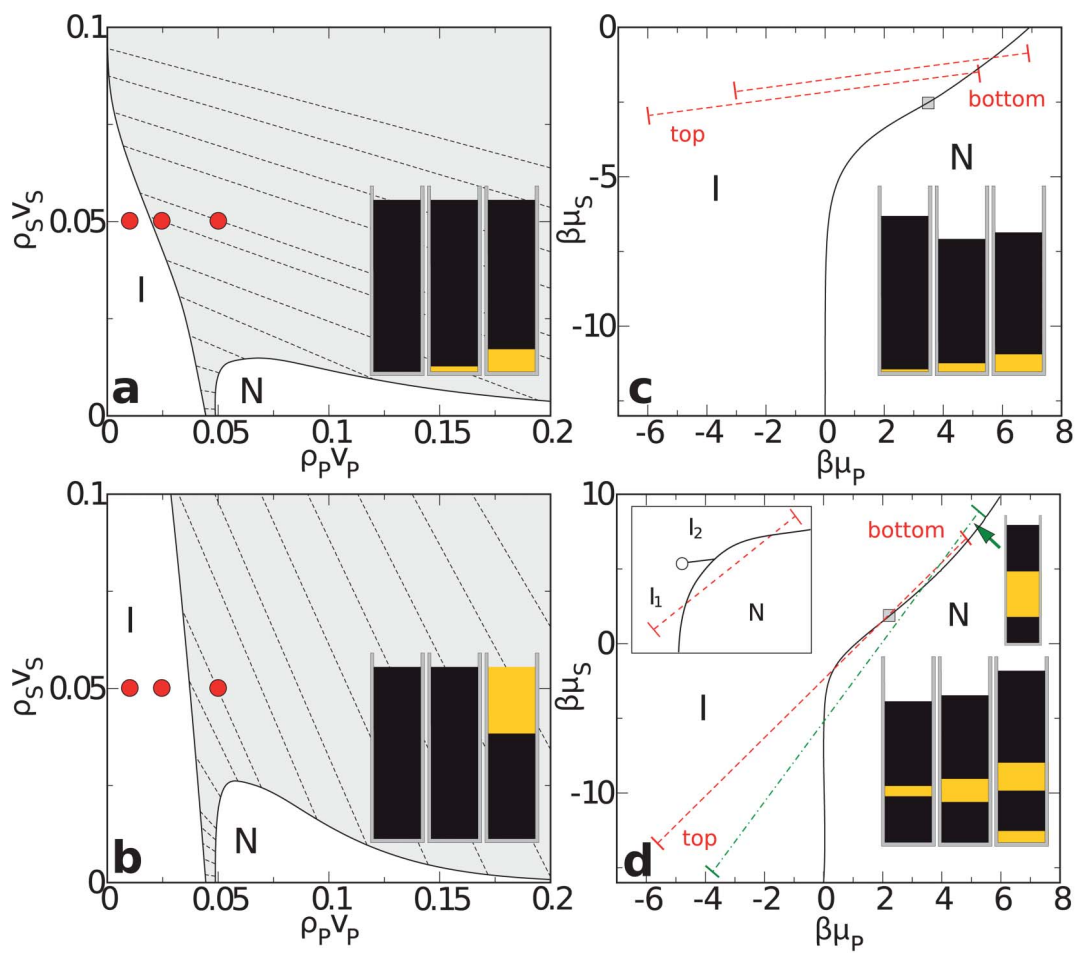

**Figure 2** | **Theoretical bulk phase diagrams of mixtures of hard spheres and infinitely thin hard platelets with aspect ratio $\sigma_P^*/\sigma_S^* = 4.9$ (a, c) and 2.33 (b, d), corresponding to the experimental mixtures with $\sigma_S = 30$ nm and 74 nm, respectively.** (a) and (b) show the plane of densities $\rho_P$, $\rho_S$, scaled by the corresponding bare particle volume $v_i$, $i = S, P$. The experimental values for average densities are indicated as statepoints (filled circles) with $\eta_S = 0.05$ and $\eta_P = 0.01, 0.025$ and 0.05. The insets illustrate the predictions for the early equilibrium of the experimental mixtures where the nematic (isotropic) regions are shown in bright yellow (black), to be compared with the experimental observations after 12 hours (second column of Fig. 1). (c) and (d) show the phase diagram in the plane of chemical potentials $\mu_P$, $\mu_S$, scaled by $\beta = 1/(k_BT)$. Sedimentation paths are indicated as red dashed lines. For clarity only the paths for $\eta_P = 0.01$ (lower) and 0.05 (upper) in (c) and that for $\eta_P = 0.025$ in (d) are shown. The square indicates the point of inflection on the binodal. The slope of the binodal in (c, d) is related to the slope of the tielines in (a, b) via[21] $d\mu_S/d\mu_P|_{coex} = -\Delta\rho_P/\Delta\rho_S$; here $\Delta\rho_i$ is the density jump of species $i$. The cartoons in (c, d) are to be compared with the experimental observations after 20 months (fourth column of Fig. 1). Also shown is a path for $\xi_P = 4$ mm and $\eta_P = 0.05$ (green dot-dashed line and upper cartoon). The inset in (d) shows a sedimentation path for a floating nematic in a schematic phase diagram with an IIN triple point and II demixing that ends in a critical point (circle).

As $\xi_i \gg \sigma_i$, we trust the local density approximation[20], and obtain the density profiles simply via $\rho_i(z) = \rho_i^{eq}(\mu_S(z), \mu_P(z))$. When taking $z$ as a parameter, Eq. (1) represents a line with slope $\xi_P/\xi_S$ in the $\mu_P$, $\mu_S$-plane. Its endpoints are determined by $H$ (the position of the upper meniscus of the dispersion) and the constraints of fixed average densities, $\int_0^H dz \rho_i(z)/H = \eta_i/v_i$.

As for the small spheres $\xi_S \gg \xi_P$, the corresponding sedimentation paths are almost horizontal, cf. Fig. 2c. Each path crosses only once from the nematic to the isotropic side of the binodal. The agreement with the experimental findings after 20 months is very good; we took partial evaporation of the solvent into account by correspondingly decreasing the value of $H$ upon $N_i =$ const. The system with $\eta_P = 0.05$ shows clear compression of the nematic layer, and that with $\eta_P = 0.01$ possesses a bottom nematic layer, both as experimentally observed.

In the mixture with larger spheres, their larger buoyant mass creates significantly steeper sedimentation paths, such that two crossings of the binodal can occur, cf. Fig. 2d. The path for $\eta_P = 0.025$ starts in the isotropic region, traverses the nematic, and re-enters the isotropic region. This implies the experimentally observed floating nematic state. For $\eta_P = 0.01$ the nematic region is just missed, and the path lies entirely in the isotropic region (not shown). However, increasing the value to 0.020, the experimental thin floating nematic can be very well reproduced (leftmost cartoon in Fig.2d).

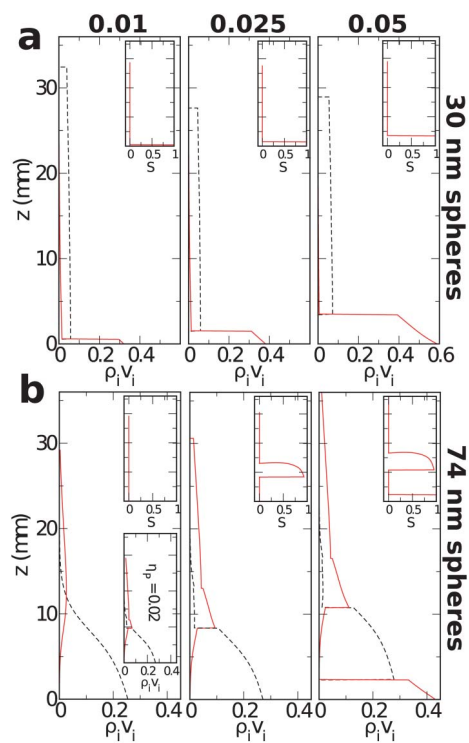

**Figure 3 | Sedimentation profiles obtained from density functional theory for mixtures of infinitely thin hard platelets and hard spheres with aspect ratio 4.9 (a), and 2.33 (b), representing the effective aspect ratio in the experimental mixtures with spheres of 30 nm and 74 nm bare diameter, respectively.** Shown are density profile for spheres $\rho_S(z)v_S$ (dashed black lines) and for platelets $\rho_P(z)v_P$ (red dashed lines) as a function of height $z$. The insets show the nematic order parameter $S(z)$ as a function of $z$. Note that the density maximum[5] of platelets in (b) for $\eta_P = 0.01$ is at the location of the experimentally observed thin floating nematic. The lower inset in (b) shows density profiles for $\eta_P = 0.020$, corresponding to the thin floating nematic in the leftmost cartoon in Fig. 2d.

The agreement for $\eta_P = 0.025$ is excellent. For $\eta_P = 0.05$ the point of inflection on the binodal facilitates an additional crossing with the sedimentation path at high values of $\mu_i$. The path already starts in the nematic region and hence an additional bottom nematic layers occurs. As equilibration is expected to be very slow at the dense bottom of the corresponding experimental sample, it is not inconceivable that such a four-layered state would develop after still longer waiting times. The slight instability of the lower IN interface in the sample with $\eta_S = 0.025$ could be a further hint at such behaviour. However, if the thick floating nematic is the equilibrium state for $\eta_P = 0.05$, it can easily arise from a slightly steeper sedimentation path, e.g. $\xi_P = 4$ mm, see Fig. 2d. The floating nematic is a robust feature, independent of whether isotropic-isotropic demixing exists or not (see inset to Fig. 2d). We present density profiles, $\rho_S(z)$ and $\rho_P(z)$, and nematic order parameter profiles, $S(z)$, in Fig. 3. The floating nematic is sandwiched between essentially a bottom dense isotropic sphere liquid and a top isotropic platelet liquid.

## Discussion

That particles can accumulate at well-defined heights under gravity (or centrifugation) is well-known[5]. In order to relate our theory, which is centered around the concept of a local chemical potential, to the "generalized Archimedes principle" by Piazza et al[5], where the chemical potential does not explicitly appear, one can use the thermodynamic identity $\partial \Pi / \partial \rho_i = \sum_j \rho_j \partial \mu_j / \partial \rho_i$, when starting from Eq. (S1) in Ref. 5 (Supplementary Material); here $\Pi$ is the pressure and the sum is over all components. Applying their theory to the low-density limit of the solute (platelet) in the sea of spheres requires a test particle calculation in order to obtain the (orientation-dependent) pair distribution function of spheres around a suspended platelet. Our present calculation avoids such computational burdens as it works directly on the level of the free energy. Note that the platelet density profile shown in the leftmost panel of Fig. 3b (i.e. for the lowest value of platelet density), is close to the true low density limit of the solute (platelet), which is the object of study in Ref. 5. This density profile is very spread out over the entire system, in striking contrast to a floating nematic layer.

A floating phase has been theoretically predicted in colloid-polymer mixtures[21]. However, the present system facilitates the experimental observation as $\xi_P/\xi_S$ can be controlled easily by the size ratio of the two species. Our considerations generalize to multi-component mixtures, as sedimentation paths will remain straight lines, cf. Eq. (1), which cut through binodal (hyper)surfaces. They also apply to atomic systems, where $\xi_i$ can be of the order of km, and to geophysical scales, where the strength of gravity depends on distance, as is relevant e.g. for the radial composition of the Earth[22]. Furthermore our results show that care is needed for the correct interpretation of phase coexistence under gravity; in particular the number of sedimentation layers can exceed the number of coexisting bulk phases allowed by the Gibbs phase rule[23].

1. Perrin, J. Mouvement brownien et molécules. *J. Phys. (Paris)* **9**, 5 (1910).
2. Biben, T., Hansen, J.-P. & Barrat, J.-L. Density profiles of concentrated colloidal suspensions in sedimentation equilibrium. *J. Chem. Phys.* **98**, 7330 (1993).
3. van der Beek, D., Schilling, T. & Lekkerkerker, H. N. W. Gravity-induced liquid crystal phase transitions of colloidal platelets. *J. Chem. Phys.* **121**, 5423 (2004).
4. Piazza, R., Bellini, T. & Degiorgio, V. Equilibrium sedimentation profiles of screened charged colloids: A test of the hard-sphere equation of state. *Phys. Rev. Lett.* **71**, 4267 (1993).
5. Piazza, R., Buzzaccaro, S., Secchi, E. & Parola, A. What buoyancy really is. A generalized archimedes' principle for sedimentation and ultracentrifugation. *Soft Matter* **8**, 3112 (2012).
6. Brambilla, G., Buzzaccaro, S., Piazza, R., Berthier, L. & Cipelletti, L. Highly nonlinear dynamics in a slowly sedimenting colloidal gel. *Phys. Rev. Lett.* **106**, 118302 (2011).
7. Wensink, H. H. & Lekkerkerker, H. N. W. Sedimentation and multi-phase equilibria in mixtures of platelets and ideal polymer. *Europhys. Lett.* **66**, 125 (2004).
8. van der Kooij, F. M. & Lekkerkerker, H. N. W. Formation of nematic liquid crystals in suspensions of hard colloidal platelets. *J. Phys. Chem. B* **102**, 7829 (1998).
9. van der Kooij, F. M., Kassapidou, K. & Lekkerkerker, H. N. W. Liquid crystal phase transitions in suspensions of polydisperse plate-like particles. *Nature* **406**, 868 (2000).
10. Vroege, G. J., Thies-Weesie, D. M. E., Petukhov, A. V., Lemaire, B. J. & Davidson, P. Smectic liquid-crystalline order in suspensions of highly polydisperse goethite nanorods. *Adv. Mat.* **18**, 2565 (2006).
11. Adams, M., Dogic, Z., Keller, S. L. & Fraden, S. Entropically driven microphase transitions in mixtures of colloidal rods and spheres. *Nature* **393**, 349 (1998).
12. van der Beek, D. *et al.* Isotropic-nematic interface and wetting in suspensions of colloidal platelets. *Phys. Rev. Lett.* **97**, 087801 (2006).
13. Doshi, N. *et al.* Structure of colloidal sphere-plate mixtures. *J. Phys.: Condens. Matt.* **23**, 194109 (2011).
14. Esztermann, A., Reich, H. & Schmidt, M. Density functional theory for colloidal mixtures of hard platelets, rods, and spheres. *Phys. Rev. E* **73**, 011409 (2006).
15. Rosenfeld, Y. Free-energy model for the inhomogeneous hard-sphere fluid mixture and densityfunctional theory of freezing. *Phys. Rev. Lett.* **63**, 980–983 (1989).
16. Onsager, L. The effects of shape on the interaction of colloidal particles. *Ann. N. Y. Acad. Sci.* **51**, 627 (1949).
17. de las Heras, D. & Schmidt, M. Bulk fluid phase behaviour of colloidal platelet-sphere and platelet-polymer mixtures. *Phil. Trans. A* (to appear in the issue on *New frontiers in anisotropic fluid-particle composites*), arXiv:1210.2551.
18. Eppenga, R. & Frenkel, D. Monte Carlo study of the isotropic and nematic phases of infinitely thin hard platelets. *Mol. Phys.* **2**, 1303 (1984).
19. Harnau, L. Structure and thermodynamics of platelet dispersions. *Mol. Phys.* **106**, 1975 (2008).
20. Hansen, J. P. & McDonald, I. R. *Theory of Simple Liquids* (Academic Press, London, 2006), 3rd edn.
21. Schmidt, M., Dijkstra, M. & Hansen, J.-P. Floating liquid phase in sedimenting colloid-polymer mixtures. *Phys. Rev. Lett.* **93**, 088303 (2004).

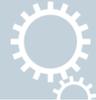

www.nature.com/scientificreports

22. Andrault, D. *et al*. Solid-liquid iron partitioning in earth's deep mantle. *Nature* **487**, 354 (2012).
23. Gibbs, J. W. (1875–1878). *On the Equilibrium of Heterogeneous Substances*, Transactions of the Connecticut Academy.

### Acknowledgements
ND was supported by the EPSRC and by Imerys Minerals Ltd. This publication was funded by the German Research Foundation (DFG) and the University of Bayreuth in the funding programme Open Access Publishing. Thomas M. Fischer is kindly acknowledged for a critical reading of the manuscript.

### Author contributions
JvD, TC, and ND designed the experiments, with the help of DG and JP. ND carried out the experiments, supervised by JvD and TC. DdlH carried out the theoretical calculations. DdlH and MS developed the theory and wrote the manuscript.

### Additional information
**Competing financial interests:** The authors declare no competing financial interests.

**License:** This work is licensed under a Creative Commons Attribution-NonCommercial-NoDerivative Works 3.0 Unported License. To view a copy of this license, visit http://creativecommons.org/licenses/by-nc-nd/3.0/

**How to cite this article:** de las Heras, D. *et al*. Floating nematic phase in colloidal platelet-sphere mixtures. *Sci. Rep.* **2**, 789; DOI:10.1038/srep00789 (2012).